\documentclass{article}

\usepackage[preprint]{neurips_2026}

\usepackage[utf8]{inputenc} 
\usepackage[T1]{fontenc}    
\usepackage{hyperref}       
\usepackage{url}            
\usepackage{booktabs}       
\usepackage{amsfonts}       
\usepackage{nicefrac}       
\usepackage{microtype}      
\usepackage{xcolor}         

\usepackage{newtxtext}       %
\usepackage[varvw]{newtxmath}       
\usepackage{amsmath,booktabs,array,hyperref}
\usepackage{tikz}
\usetikzlibrary{positioning}
\usepackage{graphicx}

\newcommand{\ind}{\mathbb{I}}

\newcommand{\keep}{\textsc{Keep}}
\newcommand{\hybrid}{\textsc{Hybrid}}
\title{When Is Additional Retrieval Worthwhile?
A Cost-Aware Study of Evidence Gain and Answer Reliability in RAG}

\author{%
  Zhouting Zhao\\
  \texttt{noora.zhaozz@gmail.com} \\
}

\begin{document}

\maketitle

\begin{abstract}
Retrieval-augmented generation (RAG) can benefit from more sophisticated retrieval, but extra retrieval may be unnecessary for many questions and gains in evidence retrieval may not translate into faithful or correct answers. We study a cost-aware, two-action retrieval policy that chooses between a dense-retrieval-plus-reranking baseline and a hybrid lexical--dense retriever. Rather than classifying baseline retrieval as simply successful or unsuccessful, we estimate the question-specific incremental evidence-retrieval benefit of hybrid retrieval using features available before the routing decision. A regularized linear model and decision threshold are selected using document-grouped cross-validation and frozen prior to held-out evaluation. On 60 previously unseen QASPER training-split articles (188 questions, 169 evidence-evaluable), the policy triggers hybrid retrieval on 46 questions (24.5\%) and obtains the same evidence Hit@10 as always-hybrid retrieval (78.70\%, versus 76.92\% for always-keep). Paired, paper-cluster bootstrap intervals do not establish a reliable utility advantage. We further compare generated answers from a 4-bit Qwen3-4B model under the three retrieval policies. Provisional answer-level assessments show no clear correctness or faithfulness improvement after partial review, despite the evidence-retrieval result. The generation assessments remain model-assisted and lack independent human validation. These results expose an important distinction between retrieving relevant evidence efficiently and improving the reliability of the generated answer. The work is presented as a reproducible empirical study, not a claim of statistically established answer-quality superiority.
\end{abstract}

\section{Introduction}

Retrieval-augmented generation (RAG) has emerged as an effective approach for improving the factual grounding of large language models (LLMs) by incorporating information from external knowledge sources \citep{lewis2020rag,gao2023survey}. By combining parametric knowledge with retrieved evidence, RAG systems can support knowledge-intensive question answering while offering greater transparency and access to domain-specific information. Nevertheless, the reliability of generated answers remains dependent on both the quality of retrieved evidence and the generator's ability to interpret and use that evidence correctly.

A central limitation of conventional RAG pipelines is their reliance on fixed retrieval strategies. Regardless of the difficulty of a question or the quality of initially retrieved evidence, the same retrieval procedure is typically applied. However, different queries may benefit from different retrieval strategies. Dense retrieval captures semantic similarity, while lexical retrieval can recover information involving specific terminology or exact expressions. Combining these approaches may improve evidence coverage, but additional candidates can also introduce irrelevant information, displace useful passages during reranking, and increase computational cost.

Recent studies have explored adaptive retrieval as a way to address these limitations. Adaptive-RAG \citep{jeong2024adaptive} selects retrieval strategies according to question complexity, while active retrieval approaches such as FLARE \citep{jiang2023flare} dynamically retrieve additional information during generation. Corrective RAG \citep{yan2024crag} introduces retrieval-quality assessment to trigger corrective actions. These studies demonstrate the value of adapting retrieval behavior, but also motivate a more specific question: whether the incremental benefit of an additional retrieval strategy can be predicted from the evidence already available.

Importantly, improvements in retrieval effectiveness do not necessarily translate into more reliable generated answers. A retriever may identify relevant evidence without providing all the information needed to answer a question. Conversely, a generator may produce an incomplete, incorrect, or unsupported answer even when sufficient evidence is present. Previous studies on attribution and factuality evaluation, including ALCE \citep{gao2023alce}, FActScore \citep{min2023factscore}, and RAGAs \citep{es2024ragas}, highlight the importance of distinguishing factual correctness from evidential support. Consequently, evaluating a retrieval strategy solely through evidence recall or hit rate may provide an incomplete picture of its downstream utility.

This paper investigates the following research question: \emph{To what extent can cost-aware adaptive retrieval improve the correctness and faithfulness of RAG-generated answers compared with fixed retrieval strategies?}

We study a deliberately constrained two-action retrieval policy. Given an initial dense-retrieval and cross-encoder-reranking result, the policy selects either \keep{}, which retains the baseline evidence, or \hybrid{}, which uses lexical--semantic retrieval followed by reranking. Rather than predicting retrieval failure directly, the policy estimates the expected incremental evidence benefit of switching to hybrid retrieval using features observable from the baseline retrieval process. The decision additionally accounts for the cost of invoking the more expensive retrieval strategy.

A distinguishing aspect of our study is the separation between the policy's training objective and its downstream evaluation. The policy is trained to predict improvements in evidence Hit@10, whereas its practical usefulness is assessed through both retrieval-level outcomes and generation-level reliability. This design allows us to examine whether improvements in retrieved evidence translate into corresponding gains in answer correctness and faithfulness.

Our contributions are threefold:
\begin{itemize}
    \item We develop an operational two-action, cost-aware adaptive retrieval policy that predicts the incremental evidence benefit of hybrid retrieval using only baseline-observable runtime features.

    \item We conduct a document-disjoint, prediction-first held-out evaluation, comparing adaptive retrieval against fixed dense and hybrid baselines with paired paper-level uncertainty estimates.

    \item We investigate the relationship between evidence retrieval effectiveness and downstream answer reliability through an exploratory generation-level evaluation of correctness, faithfulness, and abstention.
\end{itemize}

Our findings highlight both the efficiency potential and the limitations of evidence-gain-based routing. While adaptive retrieval can reduce unnecessary hybrid retrieval calls, improvements in evidence-level metrics do not automatically imply improvements in generated-answer reliability.

\begin{figure}[t]
    \centering
    \includegraphics[width=\linewidth]{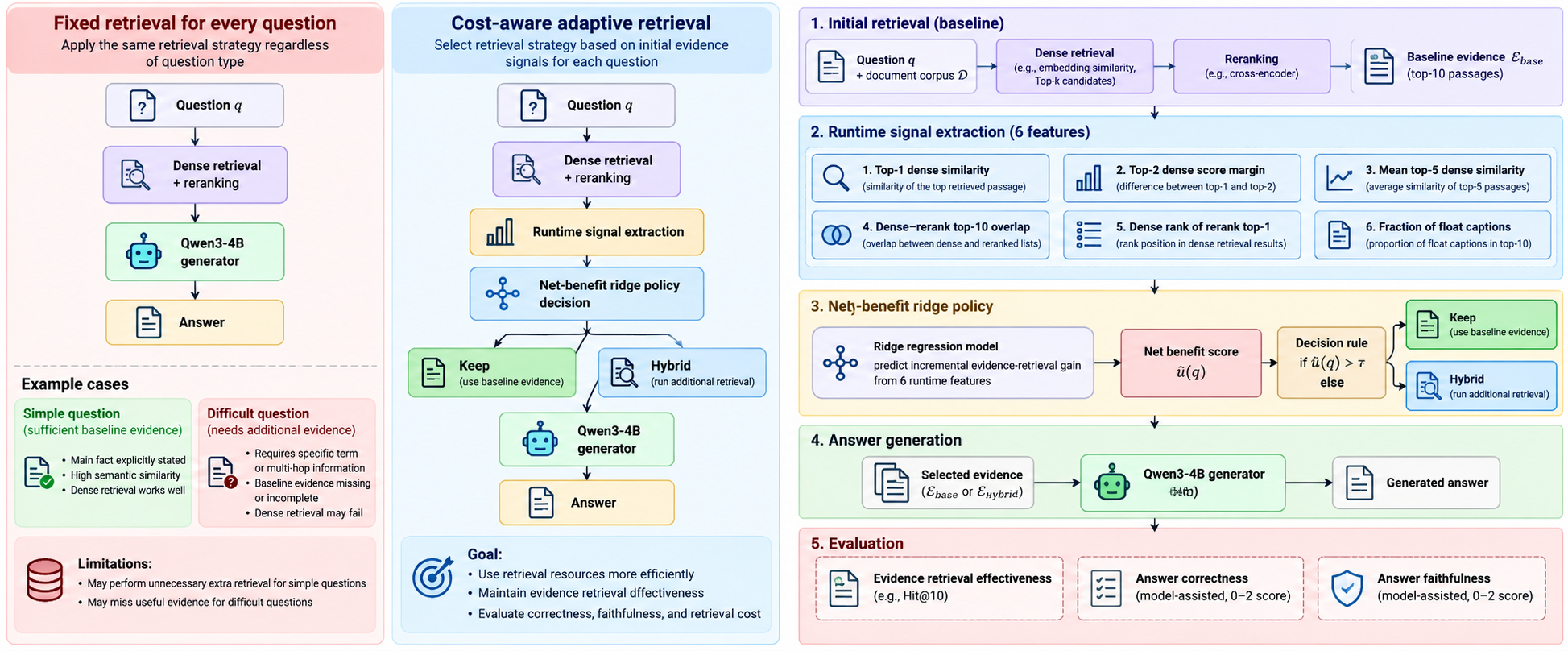}
    \caption{
    Motivation for reliability-oriented adaptive retrieval.
    Fixed retrieval strategies may provide incomplete or insufficient evidence, potentially leading to incorrect or unsupported answers.
    Adaptive retrieval uses observable baseline retrieval signals to determine whether additional hybrid retrieval is warranted.
    The central question is whether improved evidence selection can translate into more reliable generated answers at a reasonable computational cost.
    }
    \label{fig:intro_motivation}
\end{figure}
Figure~\ref{fig:intro_motivation} illustrates the motivation and overall workflow of our cost-aware adaptive retrieval approach.

\section{Related Work}

\paragraph{Retrieval-Augmented Generation.}
Retrieval-augmented generation integrates external knowledge retrieval with language-model generation to improve performance on knowledge-intensive tasks. \citet{lewis2020rag} introduced a foundational RAG framework combining parametric sequence-to-sequence models with non-parametric knowledge retrieval. Subsequent research has explored different retrieval and augmentation mechanisms, including retriever adaptation and hierarchical evidence organization \citep{shi2024replug,sarthi2024raptor}. \citet{gao2023survey} provide a broader overview of these developments. Despite substantial advances in retrieval architectures, the availability of relevant evidence does not guarantee that a language model will produce a factually correct and well-grounded answer. Our work focuses on the decision of whether to invoke an additional retrieval strategy rather than proposing a new retrieval architecture.

\paragraph{Adaptive and Active Retrieval.}
Adaptive retrieval methods aim to allocate retrieval effort according to the needs of individual queries. \citet{jeong2024adaptive} proposed Adaptive-RAG,
which selects retrieval strategies according to question complexity. \citet{jiang2023flare} introduced FLARE that performs active retrieval during text generation by identifying potentially unreliable upcoming content. \citet{asai2024selfrag} proposed Self-RAG that integrates retrieval and generation with learned reflection mechanisms that support adaptive retrieval and self-evaluation. These approaches demonstrate that retrieval need not follow a uniform procedure across all queries. Our approach differs in its narrower decision setting: instead of selecting retrieval depth from question complexity or controlling retrieval during generation, we use the output of an initial retrieval procedure to estimate the marginal evidence benefit of a predefined hybrid alternative.

\paragraph{Corrective and Hybrid Retrieval.}
Corrective Retrieval-Augmented Generation (CRAG) \citep{yan2024crag} evaluates retrieved evidence and triggers corrective retrieval operations when the available information is unreliable. Hybrid retrieval provides another mechanism for improving evidence coverage by combining complementary lexical and semantic signals. However, additional retrieval is not guaranteed to improve the final evidence ranking, since newly introduced candidates may displace previously relevant passages. We therefore frame hybrid retrieval as an optional action with potentially positive, neutral, or negative incremental outcomes. Rather than treating low retrieval confidence as sufficient justification for correction, our policy estimates the expected change in evidence retrieval success.

\paragraph{RAG Reliability and Evaluation.}
Evaluating RAG systems requires distinguishing retrieval quality from generation quality. RAGAs \citep{es2024ragas} introduces a framework for assessing different components of RAG pipelines, including faithfulness and answer quality. \citet{ru2024ragchecker} introduced RAGChecker which further emphasizes fine-grained diagnosis of retrieval and generation behavior. Complementary work on factuality and attribution, such as FActScore \citep{min2023factscore} and ALCE \citep{gao2023alce}, demonstrates the importance of verifying generated claims and their supporting evidence. Motivated by these findings, we evaluate retrieval-level effectiveness separately from answer correctness and faithfulness. This distinction is particularly important because our adaptive policy is optimized using an evidence-level proxy rather than direct supervision from answer-quality labels.

Overall, our work investigates whether a lightweight, cost-aware policy trained on incremental evidence retrieval outcomes can provide a useful efficiency--reliability trade-off. We do not claim novelty in the underlying retrieval algorithms or regression model; instead, we empirically examine the relationship between marginal retrieval benefit, adaptive routing, and downstream answer reliability.

\section{Problem Formulation and Retrieval Methods}
Let $q_i$ denote a question about source article $d_i$. Both retrieval arms search only evidence units belonging to $d_i$; they do not search other articles. Each article is represented by paragraph, section-heading, and figure/table-caption text units. The figure or table body may not be available, so a caption is only a proxy for figure-dependent evidence.

The \keep{} arm embeds the question and evidence with \texttt{all-MiniLM-L6-v2}, ranks candidates by cosine similarity, and reranks the top 30 with \texttt{ms-marco-MiniLM-L6-v2} cross-encoder, returning 10 evidence units. The \hybrid{} arm combines dense ranking with BM25 lexical ranking via reciprocal rank fusion (RRF), uses a candidate pool of 30, and applies the same cross-encoder reranker:
\begin{equation}
 \operatorname{RRF}(e\mid q)=\sum_{r\in\{\mathrm{dense},\mathrm{BM25}\}:e\in r}\frac{1}{60+\operatorname{rank}_r(e)}.
\end{equation}
The two top-10 evidence sets are denoted $E_i^K$ and $E_i^H$. A retrieval label $h_i^a\in\{0,1\}$ indicates whether $E_i^a$ contains any aligned gold evidence for arm $a\in\{K,H\}$. Gold labels are used only for offline policy training and evaluation, never as deployment features.

\section{Cost-Aware Incremental-Benefit Policy}
The offline training target is the realized retrieval difference
\begin{equation}
 \Delta_i=h_i^H-h_i^K\in\{-1,0,1\}.
 \label{eq:delta}
\end{equation}
Positive values represent rescues, negative values harms, and zeros ties. The deployment-time feature vector $x_i$ has six quantities computed from the \keep{} arm alone: top-1 dense similarity, top-2 dense margin, mean top-5 dense similarity, overlap of dense and reranked top-10 candidates, the dense rank of the rerank top-1 candidate (or 11 if absent), and the fraction of baseline top-10 units that are figure/table captions.

The fitted pipeline is median imputation, standardization, and ridge regression:
\begin{equation}
 \hat\Delta_i=\hat\beta_0+\hat\beta^\top \tilde x_i,\qquad
 \min_{\beta_0,\beta}\sum_i(\Delta_i-\beta_0-\beta^\top\tilde x_i)^2+20\|\beta\|_2^2.
\end{equation}
A unitized extra-hybrid penalty $\lambda=0.02$ defines offline retrieval utility $u_i=h_i^{\pi}-\lambda\ind[\pi(x_i)=H]$. The deployed decision rule is
\begin{equation}
 \pi(x_i)=\begin{cases}\hybrid,&\hat\Delta_i\ge\tau,\\\keep,&\hat\Delta_i<\tau.\end{cases}
 \qquad\tau=0.0310.
 \label{eq:policy}
\end{equation}
At inference time, the policy uses features extracted solely from the baseline retrieval results. Hybrid retrieval is executed only when the predicted incremental benefit meets or exceeds the selected decision threshold.

The threshold was selected from group-held-out development predictions to maximize mean predicted-policy net utility under the specified penalty, with ties preferring fewer triggers. The fitted model and threshold were frozen before held-out gold evidence was opened. The penalty is an \emph{abstract unit cost}; it is not an observed time, token, energy, or monetary cost. Accordingly the derived utility is a proxy and must not be interpreted as measured end-to-end efficiency.

\begin{figure}[t]
\centering
\includegraphics[width=\linewidth]{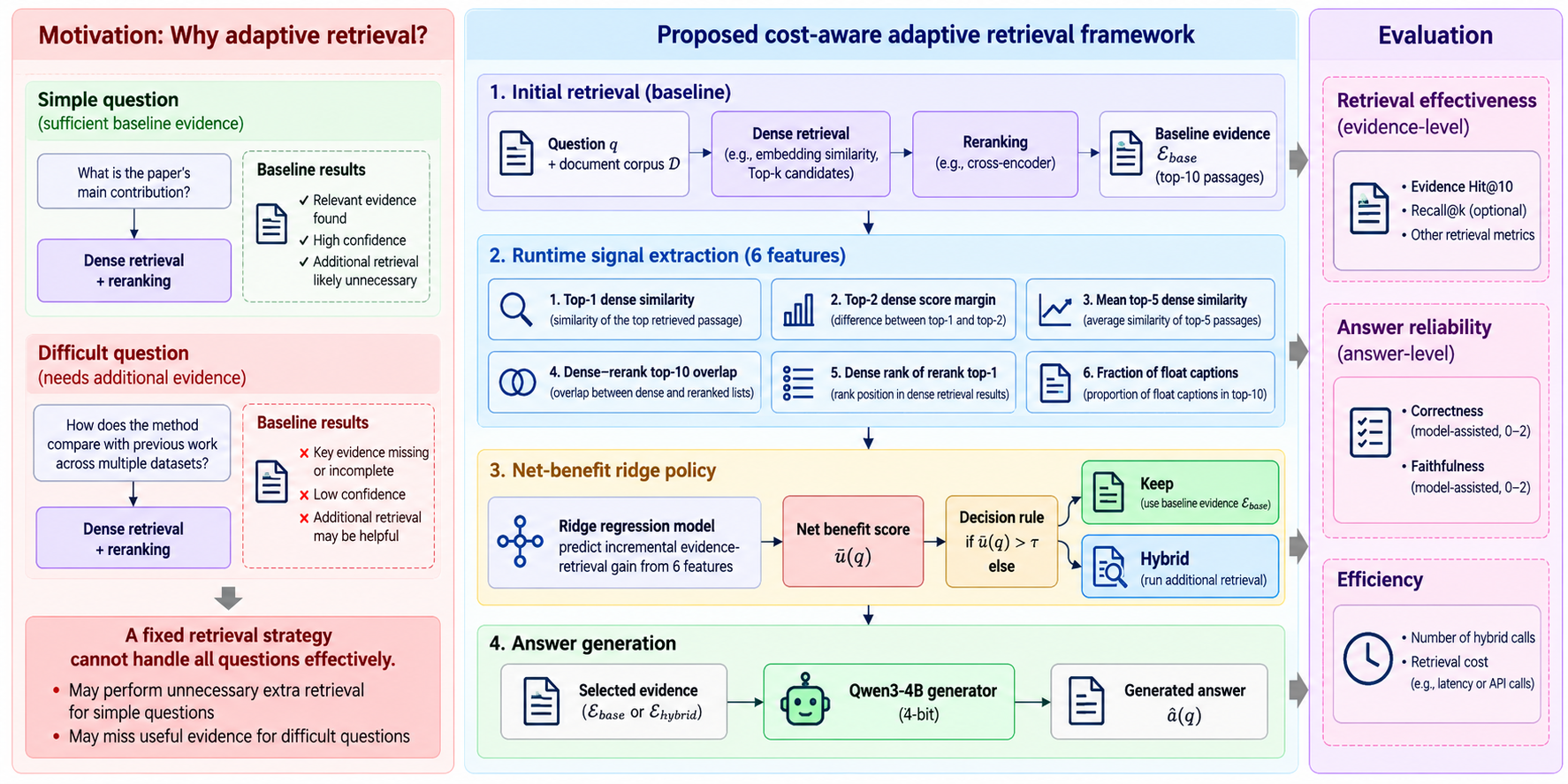}
\caption{Baseline-feature-only adaptive retrieval. The generator does not see the gold evidence or reference answers. Both retrieval alternatives are computed for offline comparison; only the routed alternative is needed operationally.}
\label{fig:pipeline}
\end{figure}

\section{Experimental Setup}
\subsection{Data and leakage controls}
The QASPER benchmark contains questions about research papers \citep{dasigi2021qasper}. Our development partition comprises 240 articles (690 questions, 599 with evidence suitable for structural retrieval evaluation). A separate 60-article partition (188 questions, 169 evidence-evaluable) was held out from policy training. Both partitions were sampled from the \emph{official QASPER training split}; the official test split was not used. In development, 16 questions were potential hybrid rescues and seven were harms, making the target highly imbalanced. All cross-validation folds group questions by article. The final training fit uses only development labels.

A separate prediction-only stage generated the held-out rankings and decisions and saved file hashes before any held-out gold evaluation. For the held-out evidence metrics, 288 referenced evidence spans could be structurally aligned; caption matches do not certify that the numerical contents of figures or tables are visible. The subsequent generation analysis on these articles is therefore an \emph{exploratory follow-up}, not another untouched test.

\subsection{Generation and evaluation protocol}
A fixed \texttt{mlx-community/Qwen3-4B-Instruct-2507-4bit} generator receives the question and top-10 evidence, with a maximum of 12,000 evidence characters and 256 output tokens. The prompt requires concise evidence-grounded answers, evidence-ID citations, and the exact string \texttt{Insufficient evidence.} when support is inadequate. We generate responses for both fixed retrieval arms. If two complete prompts are identical, their generated output is reused; the adaptive response is the previously generated response of its selected arm, not an additional independent sample. Over 188 questions, there are 376 fixed-arm answer records and 289 distinct prompts.

\textbf{Correctness} is a 0--2 assessment against QASPER reference answers: wrong/unsupported answer or unjustified abstention (0), partly correct or materially incomplete (1), and correct (2). \textbf{Faithfulness} independently scores factual claims against the evidence \emph{actually displayed to the generator}, using 0 (unsupported), 1 (partially supported), and 2 (supported). Pure abstentions have no faithfulness score and are instead analyzed by abstention appropriateness. Differences among reference annotators are flagged. These labels were provisionally assigned using the same Qwen-family judge and partially assistant-reviewed, \emph{not} independently human-validated. The score version used below is frozen for analysis but not considered definitive ground truth.

\subsection{Statistical analysis}
Policies are compared on paired questions with scores available under all three strategies. Confidence intervals for differences use 10,000 paired bootstrap resamples at the article level. Such intervals reflect article-sampling uncertainty \emph{conditional on the assigned labels}; they do not quantify the error of the language-model judge. Faithfulness is conditional on non-abstained, jointly scorable questions, so its denominator differs from correctness.

\section{Results}
\subsection{Held-out evidence retrieval}
\begin{table}[t]
\centering
\caption{Held-out evidence retrieval on 169 evaluable questions (60 articles total). Hybrid call counts are over all 188 routed questions. Utility is unitized Hit@10 minus 0.02 times the hybrid-trigger indicator.}
\label{tab:retrieval}
\begin{tabular}{lccc}
\toprule
Policy & Hit@10 (\%) & Hybrid calls & Unitized utility\\
\midrule
Always Keep & 76.92 & 0 / 188 & 0.7692\\
Always Hybrid & 78.70 & 188 / 188 & 0.7670\\
Frozen adaptive & 78.70 & 46 / 188 & \textbf{0.7819}\\
\bottomrule
\end{tabular}
\end{table}
Always-hybrid retrieval yields five potential rescues and two harms relative to keep; the frozen policy realizes four rescues and one harm, giving the same net gain in successful questions. It selects hybrid on 24.47\% of all questions (43/169 among evidence-evaluable questions), thereby avoiding 142 extra hybrid calls relative to always hybrid. The paper-level 95\% bootstrap interval for adaptive minus keep in retrieval utility is approximately $[-0.0114,0.0385]$ and for adaptive minus always hybrid is approximately $[-0.0021,0.0328]$; both include zero. Accordingly the observed result supports a reduction in hybrid-call frequency at comparable evidence hit rate, but not a statistically conclusive utility advantage. A reduction in triggered hybrid calls should not be conflated with a 75\% reduction in total runtime.

\subsection{Generated-answer correctness and faithfulness}
\begin{table}[t]
\centering
\caption{\textbf{Provisional} generation-level comparisons (scores from 0 to 2). Correctness is compared on 181 paired questions; faithfulness on 128 jointly scorable questions. Values are exploratory and not independently human validated.}
\label{tab:generation}
\begin{tabular}{lccc}
\toprule
Policy & Correctness & Faithfulness & Hybrid calls\\
\midrule
Always Keep & 1.127 & \textbf{1.844} & 0\\
Always Hybrid & \textbf{1.133} & 1.813 & 188\\
Frozen adaptive & 1.127 & 1.820 & 46\\
\bottomrule
\end{tabular}
\end{table}
Following partial manual-assisted review of high-priority correctness cases, the adaptive and keep correctness means coincide. The paired adaptive-minus-keep correctness difference is 0.000 on the 0--2 scale (95\% paper-cluster interval approximately $[-0.029,0.024]$). This result differs from an earlier, unreviewed judge-only comparison, in which the adaptive advantage was small and positive, demonstrating the sensitivity of small policy differences to adjudication. Faithfulness does not exhibit a consistent adaptive improvement over keep; in the provisional common subset, it is slightly lower. No claim of answer-reliability superiority is supported by these assessments.

The evidence and generation analyses measure different outcomes. Retrieval Hit@10 only asks whether \emph{at least one} aligned reference evidence unit appears within the top 10. A generator can still omit required facts, answer incorrectly despite sufficient evidence, or abstain. Conversely, an answer may be grounded in available evidence without satisfying a more detailed gold response. Figure/table captions may create another mismatch when gold answers require numeric values that are not present in the caption text.

\subsection{Observed computational work}
The held-out generation experiment executed 289 unique prompts (376 fixed-arm records) and recorded 1,814.409 seconds of aggregate model-generation time, excluding model loading and shared overhead. This is a workload total, not a per-policy end-to-end latency comparison. The 46 versus 188 hybrid-trigger count is the most directly validated resource-use difference for the selected policy. Without consistently measured incremental retrieval latency, hardware-resource consumption, and per-policy generation timing, a time-based cost--quality Pareto curve remains an open measurement rather than a demonstrated result.

\subsection{Separate engineering pilot: implementation diagnostics}
To support inspection of the implemented system, we additionally conducted a \emph{separate, small engineering pilot} using ten selected questions about one research PDF. This pilot is not part of the held-out QASPER evaluation above, does not use its denominators or annotation scale, and was not used to refit the frozen retrieval policy. The local FastAPI implementation compares dense retrieval without reranking (Dense Raw), Always Keep, Always Hybrid, and the frozen adaptive router, using the same Qwen3-4B model family with a 512-token output cap. The 40 saved responses received answer-level correctness judgments (pass, fail, or uncertain), with labels reused only when a response matched an already reviewed answer exactly and its hash was verified.

\begin{table}[t]
\centering
\caption{Descriptive results from the separate 10-question, single-document engineering pilot; these are \emph{not} QASPER held-out results or statistical estimates of population performance. Generated tokens exclude retrieval, reranking, and other compute costs.}
\label{tab:engineeringpilot}
\begin{tabular}{lrrrr}
\toprule
Policy & Pass & Fail & Uncertain & Generated tokens\\
\midrule
Dense Raw & 7 & 2 & 1 & 1,912\\
Always Keep & 7 & 2 & 1 & 1,672\\
Always Hybrid & 6 & 3 & 1 & 1,752\\
Frozen adaptive & 6 & 3 & 1 & 1,755\\
\bottomrule
\end{tabular}
\end{table}

The pilot did not show a correctness advantage for adaptive routing. Its purpose is diagnostic: in one case, dense retrieval retained evidence for the formula combining two fixed pretrained models whereas reranked evidence did not; in another, comparative claims about a fairness baseline lacked support; in a third, answers from all four strategies reproduced an incorrect mathematical expression despite related retrieved text. These cases are illustrative rather than causal estimates. Citation-identifier validity alone does not establish claim-level faithfulness, and one abstention remained uncertain. Context truncation and output-length truncation also occurred. No comparable end-to-end latency or total-energy measurement was established across the four methods. The original research PDF and local document index are not included in the public engineering repository, so the pilot's saved outputs are auditable but not independently regenerable from that repository alone.

\section{Discussion}
Our results separate three claims often conflated in RAG efficiency studies: retrieving more appropriate evidence, deciding when to incur the cost of a stronger retrieval arm, and improving an answer's factual correctness and grounding. The first two are partly supported by the held-out evidence evaluation: a lightweight policy maintained the always-hybrid hit rate while invoking that arm much less frequently. The third is not established. The routing regressor was trained against a binary evidence-hit difference, which is not a calibrated estimate of answer correctness or faithfulness; the disconnect is therefore unsurprising in principle and informative empirically.

The small number of rescue and harm instances limits policy learning. The final ridge is deliberately simple, but its predictions may be weakly identified when almost all training differences are zero. Moreover, per-article question dependence requires clustered rather than question-independent uncertainty estimates. These considerations make conservative interpretation essential. We do not use the post-evaluation held-out results to refit the model or readjust its threshold.

\section{Limitations and Reproducibility}
\textbf{Human validation.} Generation labels were obtained with an LLM judge using the same model family as the generator. Only a targeted subset underwent assistant-assisted review; this is not an independent blinded human-annotator study. A defensible final answer-quality claim requires independent assessment, a clear disagreement protocol, and explicit analysis of judge agreement/error. \textbf{Selective faithfulness.} The faithfulness mean excludes pure abstentions; it cannot be interpreted as unconditional system-level grounding performance. \textbf{Gold disagreement.} Conflicting QASPER annotations and unclear question wording require adjudication; scores with unresolved cases are omitted from the designated paired analysis. \textbf{Evidence representation.} Image/table caption text is not an adequate substitute for actual figure/table values. \textbf{Generalizability.} The 60-article held-out evaluation belongs to QASPER's official training split, and was subsequently inspected for exploratory generation analysis. The official test set and any independent second benchmark are untested. \textbf{Costs.} A unitized penalty and hybrid-call count are not real-time latency or token-cost measurements. Finally, the available evidence does not demonstrate strong statistically conclusive downstream correctness or faithfulness improvements.

For auditability, retained artifacts include development and held-out question/document identifiers, input and model hashes, frozen feature order and threshold, per-arm evidence rankings, immutable generated prompts, generator outputs, provisional answer labels, and article-level paired bootstrap tables.

\paragraph{Code availability.}
An engineering-oriented implementation, the frozen Ridge policy artifact, tests, saved pilot-generation results, and review/verification artifacts are available at \url{https://github.com/noorazhaoz/TrustRAG}. The repository is a curated software release rather than a complete archive of all QASPER experiments. In particular, it does not bundle the original held-out QASPER corpus, all frozen historical model inputs and outputs, or the single-document pilot PDF/index. Accordingly, the main reported QASPER statistics should not be described as end-to-end reproducible from the public repository alone. The public code and pilot artifacts facilitate implementation inspection and partial audit; independent reproduction of every result requires the corresponding data, artifact versions, and runtime environment.

\section{Conclusion}
We evaluated a cost-aware, baseline-feature-only policy for choosing between dense-plus-reranked and hybrid-plus-reranked evidence in scientific-document RAG. On a held-out set of QASPER articles, the policy matched always-hybrid evidence Hit@10 while selecting hybrid for fewer than one quarter of questions. However, exploratory generated-answer evaluations did not establish a corresponding correctness or faithfulness advantage, and the quality labels remain insufficiently independently validated. The main empirical lesson is therefore a distinction, not a victory claim: efficient evidence selection is possible in this setting, but better retrieval metrics cannot be assumed to imply more reliable generated answers.

\bibliographystyle{plainnat}
\bibliography{bib}



\end{document}